\documentclass[12pt]{article}
\usepackage{graphicx}

\begin{document}

\title{Effective mass splitting of neutron and proton and isospin emission in heavy-ion collisions}

\author{Zhao-Qing Feng \footnote{Corresponding author. Tel. +86 931 4969215. \newline \emph{E-mail address:} fengzhq@impcas.ac.cn (Z.-Q. Feng)}}
\date{}
\maketitle

\begin{center}
\small \emph{Institute of Modern Physics, Chinese Academy of Sciences, Lanzhou 730000, People's Republic of China}
\end{center}

\textbf{Abstract}
\par
Within the framework of an isospin and momentum dependent transport model, the emissions of isospin particles (nucleons and light clusters) squeezed out in heavy-ion collisions are investigated as probes of the poorly known symmetry energy at high baryon density. Two different mass splittings of neutrons and protons in nuclear medium as $m_{n}^{\ast}>m_{p}^{\ast}$ and $m_{n}^{\ast}<m_{p}^{\ast}$ are used in the model and their influence on the isospin emission in heavy-ion collisions is discussed thoroughly. The competition between the stiffness and the momentum dependence of the symmetry potential on reaction dynamics are compared and systematically analyzed. It is found that the difference of the neutron and proton directed flows and the transverse momentum distribution of the neutron/proton ratio are sensitive to the stiffness of the symmetry energy, which can not be changed with the controversial effective mass splitting. The elliptic flows of free nucleons at high transverse momentum within mid-rapidity emission are a promising observable as distinguishing the nucleon effective mass splitting.

\bigskip

It has been well established the fact that the effective nucleon mass in nuclear matter or finite nuclei deviates from its vacuum value \cite{Je76,Ma85}. Moreover, a splitting of neutron and proton effective mass (here, the nonrelativistic mass (Landau mass) is concerned in accordance with the definition in Ref. \cite{Da05}) appears in neutron-rich matter, which increases with the isospin asymmetry. Predictions of the mass splitting based on nuclear many-body theories differ widely. Based on the realistic nucleon-nucleon interactions, the Brueckner-Hartree-Fock (BHF) and Dirac-Brueckner-Hartree-Fock (DBHF) calculations predict a neutron-proton mass splitting of $m^{\ast}_{n}>m^{\ast}_{p}$ \cite{Da05,Zu99,Fr02}. However, the relativistic mean-field (RMF) theory gives a contrary result after the inclusion of the isovector mesons $\rho$ and $\delta$ \cite{Li02,Ba05}. The Skyrme-Hartree-Fock (SHF) predicts both splittings of $m^{\ast}_{n}>m^{\ast}_{p}$ and $m^{\ast}_{n}<m^{\ast}_{p}$ exist with different Skyrme parameters \cite{Ba05}. More realistically, the mass splitting of neutron and proton has been constrained from the energy dependence of the Lane (symmetry) potential by the nucleon-nucleus scattering experimental data, which moderately supports the mass splitting of $m^{\ast}_{n}>m^{\ast}_{p}$ \cite{Li04}. Further constraints of the effective mass splitting from heavy-ion collisions are still necessary and its influence on reaction dynamics would be interesting. The Landau mass splitting of neutron and proton in neutron-rich matter results from the momentum dependence of the symmetry potential, which directly affects the isospin transport in heavy-ion collisions and consequently the extraction of the density dependence of the symmetry energy.

The exact knowledge of the density dependence of the nuclear symmetry energy is still poorly known, in particular at high baryon densities, but which has important application in nuclear physics itself and also astrophysics. Heavy-ion reactions with neutron-rich beams provide an unique opportunity to explore the density-dependent symmetry energy in a broad domain of density. A symmetry energy of the form $E_{\textrm{sym}}(\rho)\approx 31.6(\rho/\rho_{0})^{\gamma}$ MeV with $\gamma \approx 0.69-1.05$ has been roughly extracted for densities between 0.1$\rho_{0}$ and 1.2$\rho_{0}$ (where $\rho_{0}$ is the nuclear saturation density) \cite{Li08}. However, the high-density behavior of nuclear symmetry energy predicted by transport models associated with the existing experimental data deviates largely because of the inconsistent treatment of the mean-field potentials and the unclear in-medium nuclear interaction and properties of resonances \cite{Fe06,Xi09,Fe10a,Ru11}, and an opposite conclusion was drawn. Furthermore, new experimental data related to high-density observables and modifications of transport models including the in-medium effects in two-body collisions and in mean-field propagation, are very anticipated. In this work, we present systematic investigations of the effective mass splitting of neutron and proton and its influence on reaction dynamics. An extraction of the symmetry energy from the preequilibrium nucleon emission is performed with an isospin and momentum dependent transport model (Lanzhou quantum molecular dynamics (LQMD)). The model has been successfully applied to treat the nuclear dynamics at near Coulomb barrier energies and also to describe the capture of two heavy colliding nuclides to form a superheavy nucleus \cite{Fe05,Fe08}. Further improvements of the LQMD model have been performed in order to investigate the dynamics of pion and strangeness productions in heavy-ion collisions and to extract the information of isospin asymmetric EoS at supra-saturation densities \cite{Fe10a,Fe09,Fe10b,Fe10c,Fe11a}. The momentum dependence of the symmetry potential was also included in the model, which results in a splitting of proton and neutron effective mass in nuclear medium \cite{Fe11b}.

In the LQMD model, the time evolutions of the baryons (nucleons and resonances ($\Delta$(1232), N*(1440), N*(1535))) and mesons in the system under the self-consistently generated mean-field are governed by Hamilton's equations of motion, which read as
\begin{eqnarray}
\dot{\mathbf{p}}_{i}=-\frac{\partial H}{\partial\mathbf{r}_{i}},
\quad \dot{\mathbf{r}}_{i}=\frac{\partial H}{\partial\mathbf{p}_{i}}.
\end{eqnarray}
Here we only consider the Coulomb interaction for charged hyperons. The Hamiltonian of baryons consists of the relativistic energy, the effective interaction potential and the momentum dependent interaction. The effective interaction potential is composed of the Coulomb interaction and the local potential
\begin{equation}
U_{int}=U_{Coul}+U_{loc}.
\end{equation}
The Coulomb interaction potential is written as
\begin{equation}
U_{Coul}=\frac{1}{2}\sum_{i,j,j\neq
i}\frac{e_{i}e_{j}}{r_{ij}}erf(r_{ij}/\sqrt{4L})
\end{equation}
where the $e_{j}$ is the charged number including protons and charged resonances. The $r_{ij}=|\mathbf{r}_{i}-\mathbf{r}_{j}|$ is the relative distance of two charged particles.

The local interaction potential is derived directly from the Skyrme energy-density functional and expressed as
\begin{equation}
U_{loc}=\int V_{loc}(\rho(\mathbf{r}))d\mathbf{r}.
\end{equation}
The local potential energy-density functional reads
\begin{eqnarray}
V_{loc}(\rho)=&& \frac{\alpha}{2}\frac{\rho^{2}}{\rho_{0}}+
\frac{\beta}{1+\gamma}\frac{\rho^{1+\gamma}}{\rho_{0}^{\gamma}} + g_{\tau}\rho^{8/3}/\rho_{0}^{5/3} +
\frac{g_{sur}}{2\rho_{0}}(\nabla\rho)^{2} + \frac{g_{sur}^{iso}}{2\rho_{0}}[\nabla(\rho_{n}-\rho_{p})]^{2}    \nonumber \\
&& + E_{sym}^{loc}(\rho)\rho\delta^{2},
\end{eqnarray}
where the $\rho_{n}$, $\rho_{p}$ and $\rho=\rho_{n}+\rho_{p}$ are the neutron, proton and total densities, respectively, and the $\delta=(\rho_{n}-\rho_{p})/(\rho_{n}+\rho_{p})$ is the isospin asymmetry. The coefficients $\alpha$, $\beta$, $\gamma$, $g_{\tau}$, $g_{sur}$, $g_{sur}^{iso}$ are related to the Skyrme parameters $t_{0}, t_{1}, t_{2}, t_{3}$ and $x_{0}, x_{1}, x_{2}, x_{3}$ \cite{Fe08} and the parameter Sly6 is taken in the calculation. The bulk parameters $\alpha$, $\beta$ and $\gamma$ are readjusted after inclusion the momentum term in order to reproduce the compression modulus of symmetric nuclear matter (here, K=230 MeV) and the binding energy of isospin symmetric nuclear matter at saturation density, which have the values of -296.6 MeV, 197 MeV and 1.143, respectively. The $E_{sym}^{loc}$ is the local part of the symmetry energy, which can be adjusted to mimic predictions of the symmetry energy calculated by microscopical or phenomenological many-body theories and has two-type forms as follows:
\begin{equation}
E_{sym}^{loc}(\rho)=\frac{1}{2}C_{sym}(\rho/\rho_{0})^{\gamma_{s}},
\end{equation}
and
\begin{equation}
E_{sym}^{loc}(\rho)=a_{sym}(\rho/\rho_{0})+b_{sym}(\rho/\rho_{0})^{2}.
\end{equation}
The parameters $C_{sym}$, $a_{sym}$ and $b_{sym}$ are taken as the values of 52.5 MeV, 43 MeV, -16.75 MeV and 23.52 MeV, 32.41 MeV, -20.65 MeV corresponding to the mass splittings of $m_{n}^{\ast}>m_{p}^{\ast}$ and $m_{n}^{\ast}<m_{p}^{\ast}$, respectively. The values of $\gamma_{s}$=0.5, 1., 2. have the soft, linear and hard symmetry energy with baryon density, respectively, and the Eq. (7) gives a supersoft symmetry energy, which cover the largely uncertain of nuclear symmetry energy, particularly at the supra-saturation densities. All cases cross at saturation density with the value of 31.5 MeV. The local part of the symmetry energy can be adjusted to reflect the largely uncertain behavior of the symmetry energy at sub- and supra-normal densities.

A Skyrme-type momentum-dependent potential is used in the model as follows:
\begin{eqnarray}
U_{mom}=&& \frac{1}{2\rho_{0}}\sum_{i,j,j\neq i}\sum_{\tau,\tau'}C_{\tau,\tau'}\delta_{\tau,\tau_{i}}\delta_{\tau',\tau_{j}}\int\int\int d \textbf{p}d\textbf{p}'d\textbf{r} f_{i}(\textbf{r},\textbf{p},t)  \nonumber \\
&& \times [\ln(\epsilon(\textbf{p}-\textbf{p}')^{2}+1)]^{2} f_{j}(\textbf{r},\textbf{p}',t).
\end{eqnarray}
The term is also given from the energy-density functional in nuclear matter,
\begin{eqnarray}
U_{mom}=&& \frac{1}{2\rho_{0}}\sum_{\tau,\tau'}C_{\tau,\tau'} \int\int\int d \textbf{p}d\textbf{p}'d\textbf{r} f_{\tau}(\textbf{r},\textbf{p}) [\ln(\epsilon(\textbf{p}-\textbf{p}')^{2}+1)]^{2} f_{\tau'}(\textbf{r},\textbf{p}').
\end{eqnarray}
Here $C_{\tau,\tau}=C_{mom}(1+x)$, $C_{\tau,\tau'}=C_{mom}(1-x)$ ($\tau\neq\tau'$) and the isospin symbols $\tau$($\tau'$) represent proton or neutron. The sign of $x$ determines different mass splitting of proton and neutron in nuclear medium, e.g., positive signs corresponding to the case of $m^{\ast}_{n}<m^{\ast}_{p}$. The parameters $C_{mom}$ and $\epsilon$ was determined by fitting the real part of optical potential as a function of incident energy from the proton-nucleus elastic scattering data. In the calculation, we take the values of 1.76 MeV, 500 c$^{2}$/GeV$^{2}$ for the $C_{mom}$ and $\epsilon$, respectively, which result in the effective mass $m^{\ast}/m$=0.75 in nuclear medium at saturation density for symmetric nuclear matter. The parameter $x$ is changed as the strength of the mass splitting, and the values of -0.65 and 0.65 are respective to the cases of $m^{\ast}_{n}>m^{\ast}_{p}$ and $m^{\ast}_{n}<m^{\ast}_{p}$.

Combined Eq. (5) and Eq. (9), we get a density, isospin and momentum-dependent single-nucleon potential in nuclear matter as follows:
\begin{eqnarray}
U_{\tau}(\rho,\delta,\textbf{p})=&&  \alpha\frac{\rho}{\rho_{0}}+\beta\frac{\rho^{\gamma}}{\rho_{0}^{\gamma}}+\frac{8}{3}g_{\tau}\rho^{5/3}/\rho_{0}^{5/3}+
E_{sym}^{loc}(\rho)\delta^{2} + \frac{\partial E_{sym}^{loc}(\rho)}{\partial\rho}\rho\delta^{2} +
E_{sym}^{loc}(\rho)\rho\frac{\partial\delta^{2}}{\partial\rho_{\tau}}          \nonumber \\
&&  + \frac{1}{\rho_{0}}C_{\tau,\tau} \int d\textbf{p}' f_{\tau}(\textbf{r},\textbf{p})[\ln(\epsilon(\textbf{p}-\textbf{p}')^{2}+1)]^{2}         \nonumber \\
&&  + \frac{1}{\rho_{0}}C_{\tau,\tau'} \int d\textbf{p}' f_{\tau'}(\textbf{r},\textbf{p})[\ln(\epsilon(\textbf{p}-\textbf{p}')^{2}+1)]^{2}.
\end{eqnarray}
Here $\tau\neq\tau'$, $\partial\delta^{2}/\partial\rho_{n}=4\delta\rho_{p}/\rho^{2}$ and $\partial\delta^{2}/\partial\rho_{p}=-4\delta\rho_{n}/\rho^{2}$. The nucleon effective (Landau) mass in nuclear matter of isospin asymmetry $\delta=(\rho_{n}-\rho_{p})/(\rho_{n}+\rho_{p})$ with $\rho_{n}$ and $\rho_{p}$ being the neutron and proton density, respectively, is calculated through the potential as $m_{\tau}^{\ast}=m_{\tau}/ \left(1+\frac{m_{\tau}}{|\textbf{p}|}|\frac{dU_{\tau}}{d\textbf{p}}|\right)$ with the free mass $m_{\tau}$ at Fermi momentum $\textbf{p}=\textbf{p}_{F}$. Therefore, the nucleon effective mass only depends on the momentum-dependent term of the nucleon optical potential. The isovector part of the optical potential, i.e., the symmetry or Lane potential \cite{La62}, can be evaluated from the expression $U_{sym}(\rho,\textbf{p})=(U_{n}(\rho,\delta,\textbf{p})-U_{p}(\rho,\delta,\textbf{p}))/2\delta$. Shown in Fig. 1 is a comparison of the momentum dependence of symmetry potential with the mass splittings of $m_{n}^{\ast}>m_{p}^{\ast}$ in the left window and $m_{n}^{\ast}<m_{p}^{\ast}$ in the right window at the baryon densities of 0.5$\rho_{0}$, 1.0$\rho_{0}$ and 2.0$\rho_{0}$, respectively. One should note an opposite variation with nucleon momentum appears for the two different mass splittings.

We have included the resonances $\Delta$(1232), $N^{\ast}$(1440), $N^{\ast}$(1535) in nucleon-nucleon (NN) collisions and their decaying channels into pions and etas. The in-medium potentials for pion and kaon (antikaon) propagations can be chosen in the model \cite{Fe10b,Fe11a}. Furthermore, strange particles are directly created by inelastic hadron-hadron collisions \cite{Fe10c}. The probability reaching a channel in a NN collision is determined by its contribution of the channel cross section to the total cross section. The choice of the channel is done randomly by the weight of the probability.

The preequilibrium nucleons in high-energy heavy-ion collisions are mostly produced during a compression stage of two colliding partners within a very short time. Therefore, the high-density information of nuclear phase diagram are expected to be extracted from the nucleons or light complex particles, which can be constrained from the longitudinal rapidity distributions and the azimuthal emissions. Shown in Fig. 2 is a comparison of transverse emission ratios of neutron/proton and $^{3}$H/$^{3}$He within the rapidity selection of $|y/y_{proj}|<$0.25 in the $^{197}$Au+$^{197}$Au reaction at the incident energy of 400\emph{A} MeV for the near central (b=1 fm) and semi-central (b=6 fm) collisions, but different mass splitting. One can see that the influence of the symmetry energy appears at low transverse momentum and the situation does not be changed with the mass splitting. A hard symmetry energy enforces a strongly repulsive force on neutrons in the high-density domain, furthermore, squeezes out more neutrons in the preequilibrium stage of dynamical evolution. From the negative contribution of the momentum-dependent interaction of $m_{n}^{\ast}>m_{p}^{\ast}$ to the potential part of the symmetry energy in Ref. \cite{Fe11b}, one notices that the momentum-dependent potential leads to an attractive force on neutrons, in particular at high densities, which reduces the $n/p$ yields at high transverse momentum. Opposite contributions take place and a flat distribution of the n/p and $^{3}$H/$^{3}$He appears in the case of $m_{n}^{\ast}<m_{p}^{\ast}$. The results are nice consistent with the stochastic mean-field (SMF) calculations \cite{Gi10}. Experimental measurements of the momentum distributions of the n/p ratios would be interesting to get the information of the effective mass splitting in nuclear medium and also the symmetry energy at supra-normal densities.

Collective flow has been verified as a nice approach to reconstruct the reaction plane and to study azimuthal correlation of the fireball formed in heavy-ion collisions. Rather systematical data has been measured to investigate the equation of state from flow analysis \cite{An05,Re07}. The flow information can be expressed as the first and second coefficients from the Fourier expansion of the azimuthal distribution $\frac{dN}{d\phi}(y,p_{t})=N_{0}(1+2V_{1}(y,p_{t})\cos(\phi)+2V_{2}(y,p_{t})\cos(2\phi))$ \cite{Ol92}, where $p_{t}=\sqrt{p_{x}^{2}+p_{y}^{2}}$ and $y$ are the transverse momentum and the longitudinal rapidity along the beam direction, respectively. The directed (transverse) flow is defined as the first coefficient and expressed as $V_{1}=\langle p_{x}/p_{t} \rangle$, which provides the information of the azimuthal anisotropy of the transverse emission. The elliptic flow $V_{2}=\langle (p_{x}/p_{t})^{2}-(p_{y}/p_{t})^{2} \rangle$ gives the competition between the in-plane ($V_{2}>$0) and out-of-plane ($V_{2}<$0, squeeze out) emissions. The brackets indicate averaging over all events in accordance with a specific class such as rapidity or transverse momentum cut. Shown in Fig. 3 is a comparison of rapidity distributions of directed and elliptic flows of free nucleons for near central (b=1 fm) and semi-central (b=6 fm) $^{197}$Au+$^{197}$Au collisions at the incident energy of 400\emph{A} MeV with the mass splitting of $m_{n}^{\ast}>m_{p}^{\ast}$ for the hard (left window) and supersoft (right window) symmetry energies, respectively. One notices that the difference of neutron and proton transverse flows is slight for the near central collisions. The elliptic flow disappears for the near central heavy-ion collisions and the influence of symmetry energy on the elliptic flow distribution is weak. We calculated the difference of neutron and proton directed flows as a function of rapidity distribution in the $^{197}$Au+$^{197}$Au collisions (b=6 fm) at the energy of 400\emph{A} MeV with the hard and supersoft symmetry energies, but for the different mass splitting as shown in Fig. 4. One can see that an anti-flow spectrum appears in both mass splittings. Moreover, the supersoft symmetry energy gives a larger flow difference, in particular around the region of projectile (target) rapidity. The phenomena can not be influenced by changing different mass splitting. The flows of fast nucleon emissions would be a nice probe of the nuclear symmetry energy.

The effect of mass splitting of neutron and proton on fast nucleon emissions is one of the main purposes in this work. Shown in Fig. 5 is a comparison of elliptic flows of free nucleons with a complete different mass splitting for the cases of hard and supersoft symmetry energies. One notices that the momentum-dependent potential with the $m_{n}^{\ast}<m_{p}^{\ast}$ enhances the out-of-plane emissions for neutrons because of its repulsive force on neutrons. Thus, the flow is reduced and even coincides with protons. Furthermore, a larger mass splitting (stronger isovector momentum-dependent interaction) increases the squeeze out neutrons and consequently results in an lower elliptic flows for neutrons \cite{Gi10}. The stiffness of the symmetry energy slightly affect the flow spectrum. Again, we compared the transverse momentum distribution of elliptic flows of proton and neutron as shown in Fig. 6 with the mid-rapidity selection of $|y/y_{proj}|<$0.25. A similar effect of the effective mass splitting with the rapidity distribution is observed. Basically, the symmetry energy does not change the spectrum structure. The elliptic flows of hydrogen-like particles have been used to extract the high-density symmetry energy \cite{Ru11}. More sensitive observable can be seen from the flow difference between neutron and proton as shown in Fig. 7. The $m_{n}^{\ast}>m_{p}^{\ast}$ case is always larger, in particular in the domain of high transverse momentum. The findings are nice consistent with the SMF calculations \cite{Gi10} although different treatment of the momentum-dependent potential.

In summary, within the transport model (LQMD) we have investigated the effect of the momentum-dependent interaction on fast nucleon emission. Two different mass splittings, i.e., $m_{n}^{\ast}>m_{p}^{\ast}$ and $m_{n}^{\ast}<m_{p}^{\ast}$, are used in the calculations. The momentum-dependent potential plays a significant role on the fast nucleon emissions. Specifically, the high-momentum neutron-proton ratio is changed, which appears more sensitive to the mass splitting than to the stiffness of the symmetry energy. The neutron/proton ratio and the directed flow difference of neutron and proton via transverse momentum distribution in the domain of mid-rapidity are nice observables for extracting the high-density symmetry energy.

The author (Z.Q. Feng) would like to thank the hospitality during his visit in LNS-INFN, Catania. Fruitful discussions with Prof. M. Di Toro and Dr. M. Colonna are acknowledged. This work was supported by the National Natural Science Foundation of China under Grant Nos 10805061 and 11175218, the Special Foundation of the President Fund and the West Doctoral Project of Chinese Academy of Sciences.

\newpage

\begin{figure}
\begin{center}
{\includegraphics*[width=1.\textwidth]{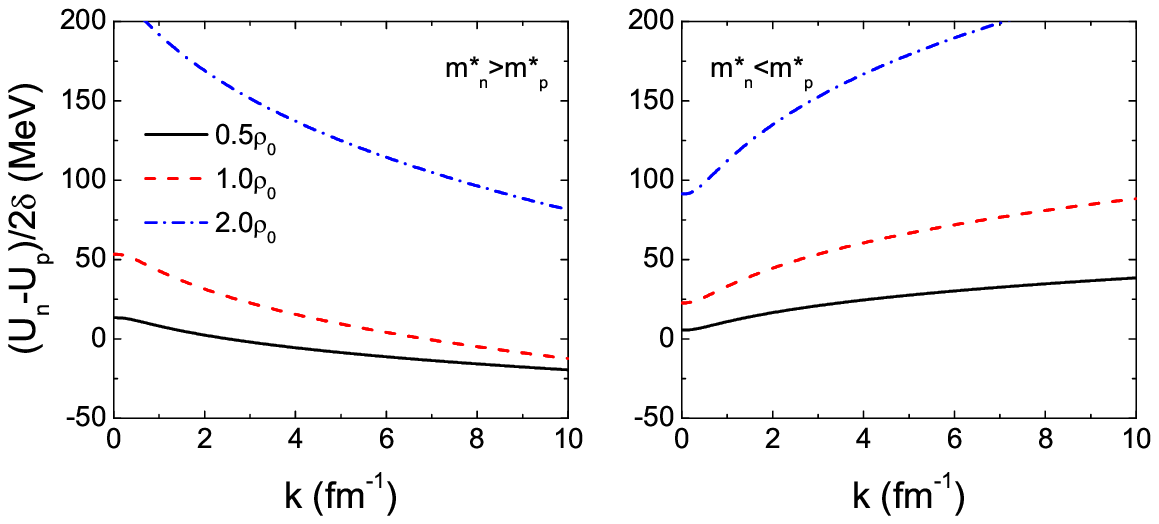}}
\end{center}
\caption{Momentum dependence of symmetry potential at baryon densities of 0.5$\rho_{0}$, 1.0$\rho_{0}$ and 2.0$\rho_{0}$ with the mass splittings of $m_{n}^{\ast}>m_{p}^{\ast}$ (left panel) and $m_{n}^{\ast}<m_{p}^{\ast}$ (right panel), respectively.}
\end{figure}

\begin{figure}
\begin{center}
{\includegraphics*[width=1.\textwidth]{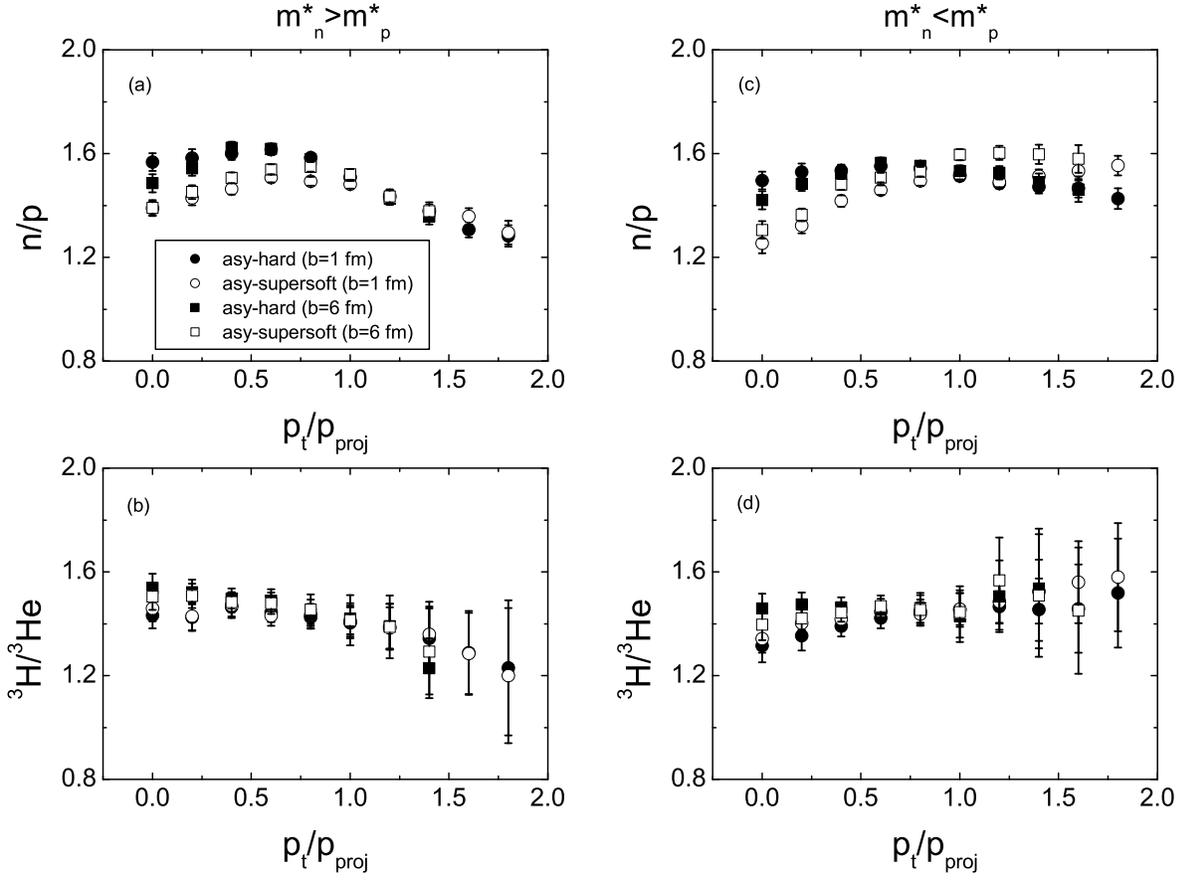}}
\end{center}
\caption{Transverse momentum distributions of neutron/proton and $^{3}$H/$^{3}$He ratios within the rapidity bin $|y/y_{proj}|<$0.25 in the $^{197}$Au+$^{197}$Au reaction at the incident energy of 400 MeV/nucleon for the near central (b=1 fm) and semi-central (b=6 fm) collisions with different mass splitting.}
\end{figure}

\begin{figure}
\begin{center}
{\includegraphics*[width=1.\textwidth]{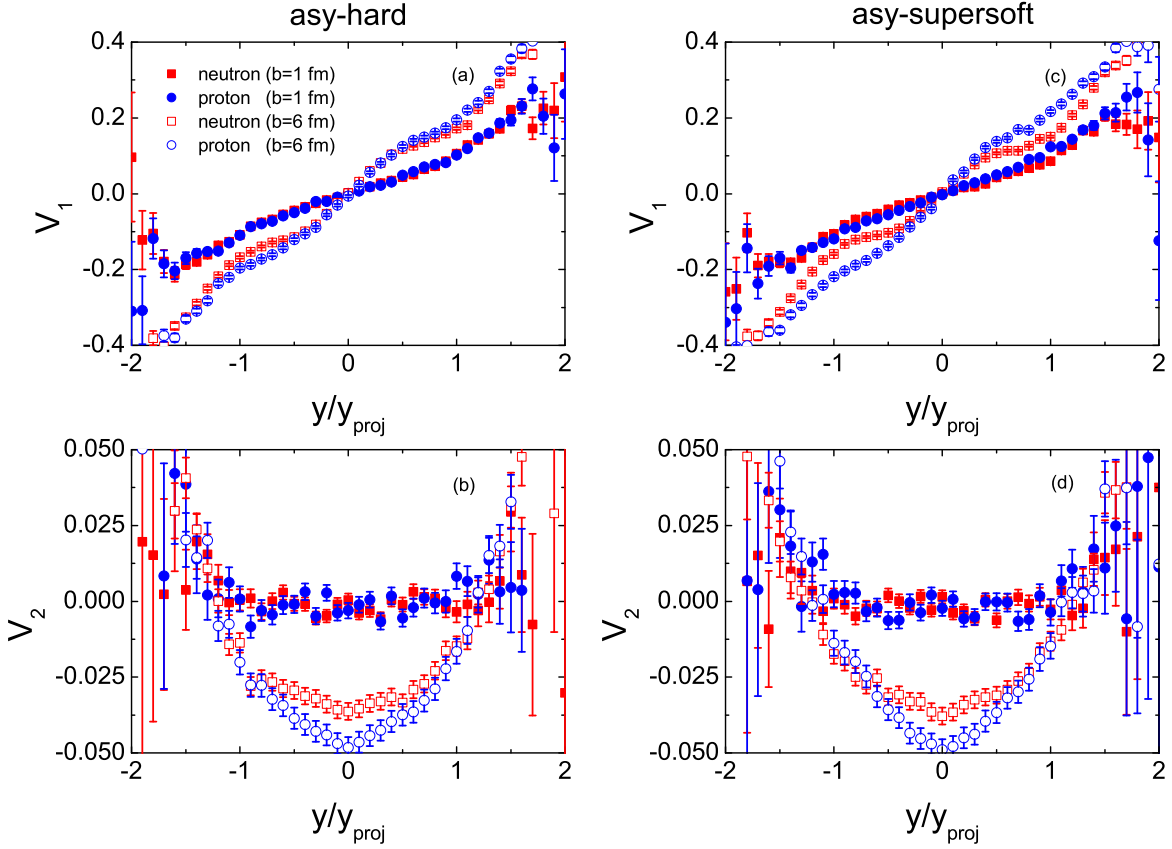}}
\end{center}
\caption{Rapidity distributions of directed and elliptic flows of free nucleons in the $^{197}$Au+$^{197}$Au reaction at the incident energy of 400 MeV/nucleon for the near central (b=1 fm) and semi-central (b=6 fm) collisions with the mass splitting of $m_{n}^{\ast}>m_{p}^{\ast}$ for the hard (left panel) and supersoft (right panel) symmetry energies, respectively.}
\end{figure}

\begin{figure}
\begin{center}
{\includegraphics*[width=1.\textwidth]{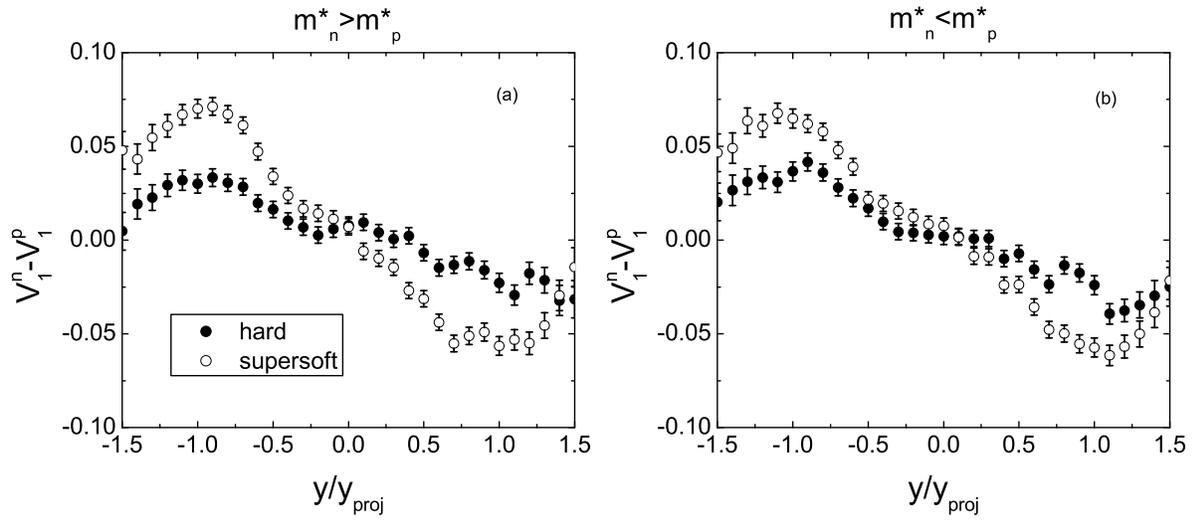}}
\end{center}
\caption{Comparison of the difference between neutron and proton directed flows in semi-central $^{197}$Au+$^{197}$Au collisions at the energy of 400 MeV/nucleon with the hard and supersoft symmetry energies for the different mass splittings of $m_{n}^{\ast}>m_{p}^{\ast}$ (left panel) and $m_{n}^{\ast}<m_{p}^{\ast}$ (right panel), respectively.}
\end{figure}

\begin{figure}
\begin{center}
{\includegraphics*[width=1.\textwidth]{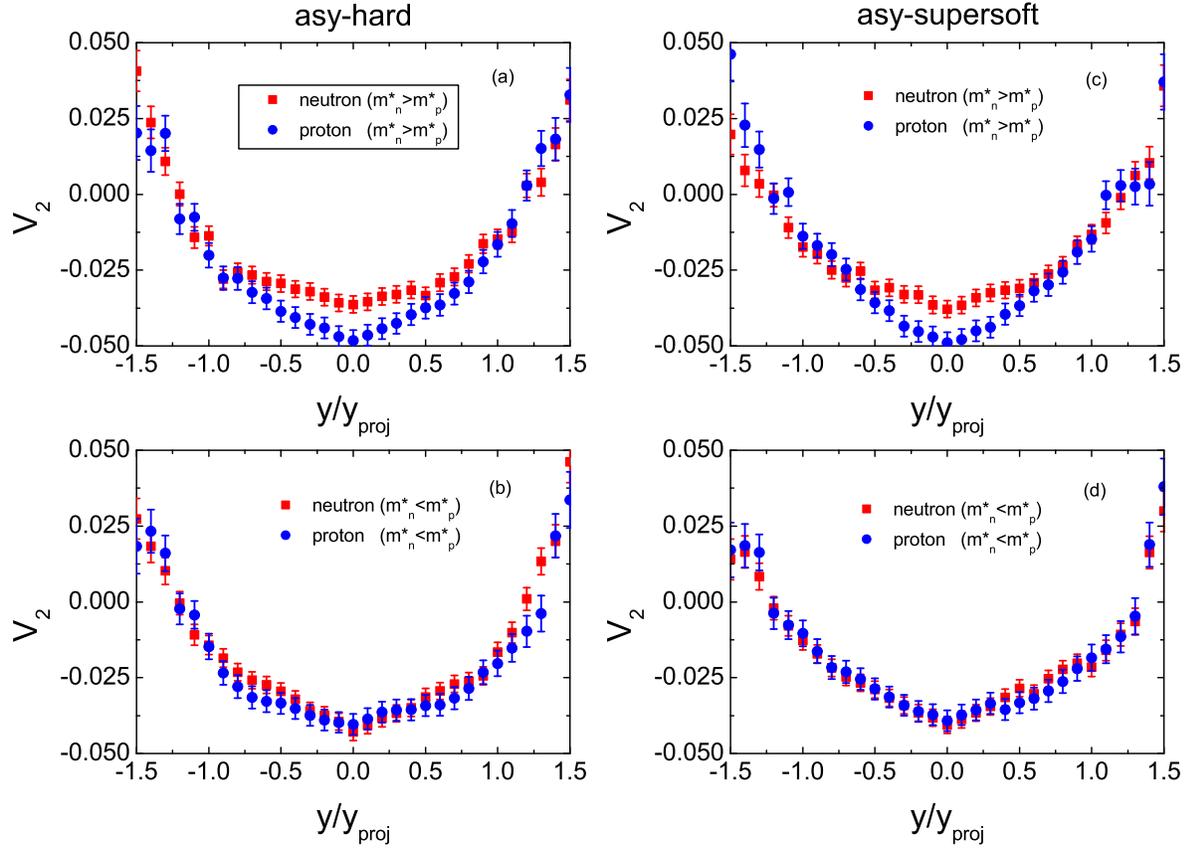}}
\end{center}
\caption{Rapidity distributions of elliptic flows of free nucleons with the different mass splittings at the cases of hard (left panel) and supersoft (right panel) symmetry energies.}
\end{figure}

\begin{figure}
\begin{center}
{\includegraphics*[width=1.\textwidth]{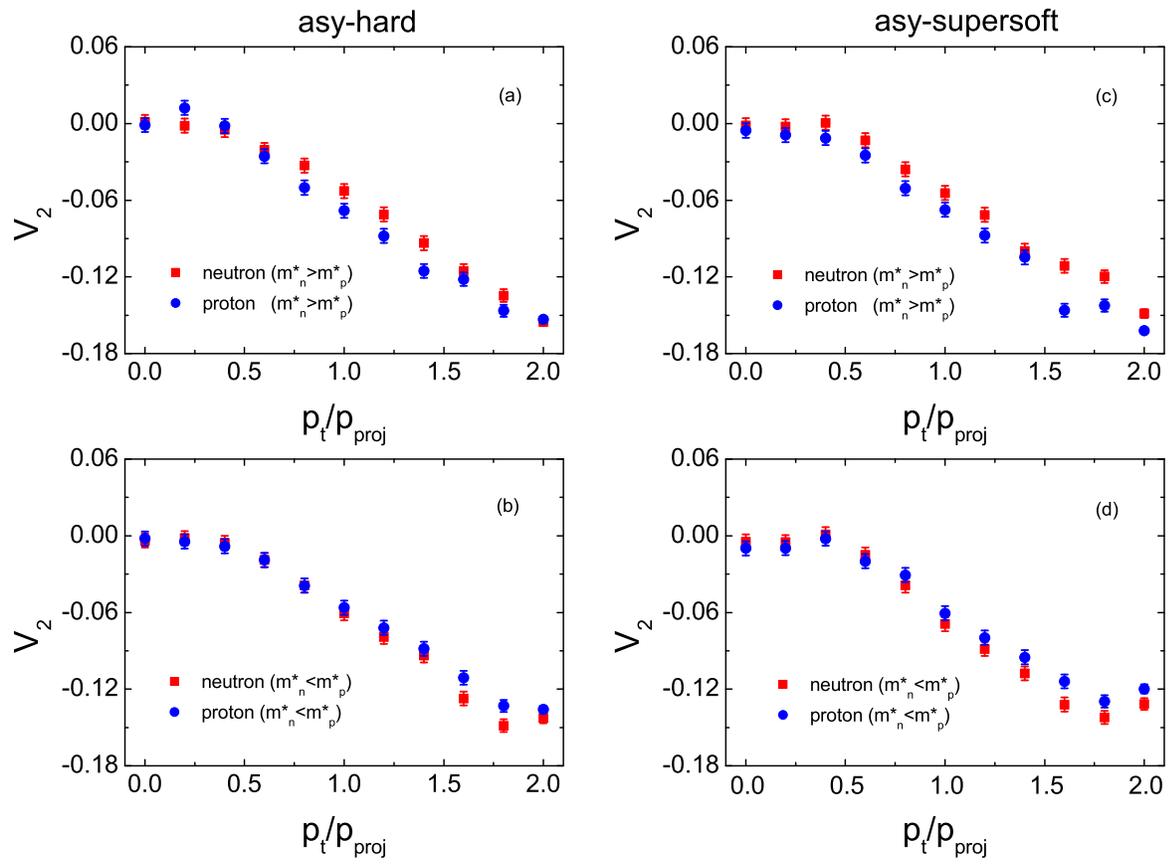}}
\end{center}
\caption{The same as in Fig. 5, but for transverse momentum distributions with the rapidity bin $|y/y_{proj}|<$0.25.}
\end{figure}

\begin{figure}
\begin{center}
{\includegraphics*[width=1.\textwidth]{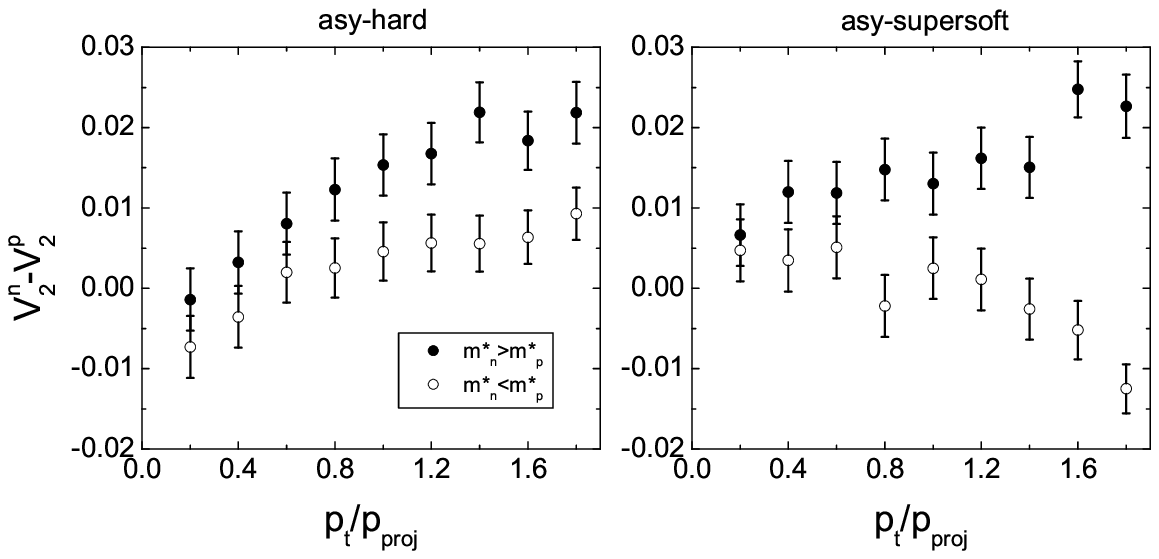}}
\end{center}
\caption{Comparison of the difference between neutron and proton elliptic flows with the rapidity bin $|y/y_{proj}|<$0.25 for the different mass splitting in the semi-central $^{197}$Au+$^{197}$Au collisions.}
\end{figure}


\begin{thebibliography}{99}

\bibitem{Je76} J.P. Jeukenne, A. Lejeune, C. Mahaux, Phys. Rep. 25 (1976) 83.
\bibitem{Ma85} C. Mahaux, P.F. Bortignon, R.A. Broglia, C.H. Dasso, Phys. Rep. 120 (1985) 1.
\bibitem{Da05} E.N.E. van Dalen, C. Fuchs, A. Faessler, Phys. Rev. Lett. 95 (2005) 022302.
\bibitem{Zu99} W. Zuo, I. Bombaci, U. Lombardo, Phys. Rev. C 60 (1999) 024605.
\bibitem{Fr02} T. Frick, Kh. Gad, H. M\"{u}ther, P. Czerski, Phys. Rev. C 65 (2002) 034321; Kh.S. A. Hassaneen, H. M\"{u}ther, Phys. Rev. C 70 (2004) 054308.
\bibitem{Li02} B. Liu, V. Greco, V. Baran, M. Colonna, M. Di Toro, Phys. Rev. C 65, 045201 (2002).
\bibitem{Ba05} V. Baran, M. Colonna, V. Greco, M. Di Toro, Phys. Rep. 410 (2005) 335.
\bibitem{Li04} B.A. Li, Phys. Rev. C 69 (2004) 064602.
\bibitem{Li08} B.A. Li, L.W. Chen, C.M. Ko, Phys. Rep. 464 (2008) 113.
\bibitem{Fe06} G. Ferini, T. Gaitanos, M. Colonna, et al., Phys. Rev. Lett. 97 (2006) 202301.
\bibitem{Xi09} Z.G. Xiao, B.A. Li, L.W. Chen, et al., Phys. Rev. Lett. 102 (2009) 062502.
\bibitem{Fe10a} Z.Q. Feng, G.M. Jin, Phys. Lett. B 683 (2010) 140.
\bibitem{Ru11} P. Russotto, P.Z. Wub, M. Zoricc, et al., Phys. Lett. B 697 (2011) 471.
\bibitem{Fe05} Z.Q. Feng, F.S. Zhang, G.M. Jin, X. Huang, Nucl. Phys. A 750 (2005) 232; Z.Q. Feng, et al., Chin. Phys. Lett. 22 (2005) 3040.
\bibitem{Fe08} Z.Q. Feng, G.M. Jin, F.S. Zhang, Nucl. Phys. A 802 (2008) 91; Z.Q. Feng, G.M. Jin, Phys. Rev. C 80 (2009) 037601.
\bibitem{Fe09} Z.Q. Feng, G.M. Jin, Chin. Phys. Lett. 26 (2009) 062501.
\bibitem{Fe10b} Z.Q. Feng, G.M. Jin, Phys. Rev. C 82 (2010) 044615.
\bibitem{Fe10c} Z.Q. Feng, G.M. Jin, Phys. Rev. C 82 (2010) 057901.
\bibitem{Fe11a} Z.Q. Feng, Phys. Rev. C 83 (2011) 067604.
\bibitem{Fe11b} Z.Q. Feng, Phys. Rev. C 84 (2011) 024610.
\bibitem{La62} A.M. Lane, Nucl. Phys. 35 (1962) 676.
\bibitem{Gi10} V. Giordano, M. Colonna, M. Di Toro, V. Greco, J. Rizzo, Phys. Rev. C 81 (2010) 044611.
\bibitem{An05} A. Andronic, et al., FOPI Collaboration, Phys. Lett. B 612 (2005) 173.
\bibitem{Re07} W. Reisdorf, et al., FOPI Collaboration, Nucl. Phys. A 781 (2007) 459; Nucl. Phys. A 848 (2010) 366.
\bibitem{Ol92} J.Y. Ollitrault, Phys. Rev. D 46 (1992) 229.

\end{thebibliography}
\end{document}